\documentclass[aps,prb,twocolumn,superscriptaddress,showpacs]{revtex4-1}
\def\be{\begin{equation}}
\def\ee{\end{equation}}
\def\bea{\begin{eqnarray}}          
\def\eea{\end{eqnarray}}
\def\bi{\begin{itemize}}
\def\ei{\end{itemize}}

\usepackage{graphicx}

\begin{document}

\title{ 
           Fermionic Projected Entangled Pair States at Finite Temperature                         
}

\author{Piotr Czarnik} 
\affiliation{Instytut Fizyki Uniwersytetu Jagiello\'nskiego
             and Centre for Complex Systems Research,
             ul. Reymonta 4, 30-059 Krak\'ow, Poland}
             
\author{Jacek Dziarmaga} 
\affiliation{Instytut Fizyki Uniwersytetu Jagiello\'nskiego
             and Centre for Complex Systems Research,
             ul. Reymonta 4, 30-059 Krak\'ow, Poland}

\date{ June 24, 2014 }

\begin{abstract}
An algorithm for imaginary time evolution of a fermionic projected entangled pair state (PEPS) 
with ancillas from infinite temperature down to a finite temperature state is presented. 
As a benchmark application, 
it is applied to spinless fermions hopping on a square lattice subject to $p$-wave pairing interactions. 
With a tiny bias
it allows to evolve the system across a high-temperature continuous symmetry-breaking phase transition.
\end{abstract}

\maketitle

%%%%%%%%%%%%%%%%%%%%%%%%%%%%%%%%%%%%%%%%%%%%%%%%%%%%%%%%%%%%%%%%%%%%%%%%%% 
\section{ Introduction } 
%%%%%%%%%%%%%%%%%%%%%%%%%%%%%%%%%%%%%%%%%%%%%%%%%%%%%%%%%%%%%%%%%%%%%%%%%%

Quantum tensor networks are a competitive tool to study strongly correlated quantum systems on a lattice. Their history 
begins with the density matrix renormalization group (DMRG) \cite{White} - an algorithm to minimize the energy of 
a matrix product state (MPS) ansatz in one dimension (1D), see Ref. \cite{Schollwoeck} for a comprehensive 
review of MPS algorithms. In the last decade MPS was generalized to a 2D ``tensor product state'' widely 
known as a projected entangled pair state (PEPS) \cite{PEPS}. Another type of tensor network is the multiscale 
entanglement renormalization ansatz (MERA) \cite{MERA}, and the branching MERA \cite{branching}, that is 
a refined version of the real space renormalization group. Being variational methods, the quantum tensor networks do not suffer 
form the notorious fermionic sign problem, and thus they can be applied to strongly correlated fermions in 2D 
\cite{fermions}. A possible breakthrough in this direction was an application of the PEPS ansatz to the t-J model \cite{PEPStJ}, 
which is a strong coupling approximation to the celebrated Hubbard Hamiltonian of the high temperature 
superconductivity \cite{highTc}. An energy of the ground state was obtained that could compete with the best 
variational Monte-Carlo results \cite{VMC}.

The tensor networks also proved to be a powerful tool to study topological spin liquids (TSL). The search for 
realistic models gained momentum after White demonstrated the spin-liquid nature of the Kagome antiferromagnet 
\cite{WhiteKagome}. This result was obtained by a {\it tour de force} application of a quasi-1D DMRG. The DMRG 
investigation of TSL's was elevated to a higher degree of sophistication in Ref. \cite{CincioVidal}. Unfortunately, 
the MPS tensor network underlying the DMRG suffers from severe limitations in two dimensions, where it can be used 
for states with a very short correlation length only. In contrast, the PEPS ansatz in Fig. \ref{FigPeps} is not 
restricted in this way. Its usefulness for TSL has already been demonstrated. In Ref. \cite{PepsRVB} it was shown 
how to represent the RVB state with the PEPS ansatz in an efficient way. In Ref. \cite{PepsKagome} PEPS was used 
to classify topologically distinct ground states of the Kagome antiferromagnet. Finally, in Ref. \cite{PepsJ1J2} 
PEPS demonstrated a TSL in the antiferromagnetic $J_1-J_2$ model.

In contrast to the ground state, finite temperature states have been explored so far mostly with the MPS 
\cite{ancillas,WhiteT}. In a way that can be easily generalized to 2D, the MPS is extended to finite 
temperature by appending each lattice site with an ancilla \cite{ancillas}. A thermal state is obtained by 
an imaginary time evolution of a pure state in the enlarged Hilbert space starting from infinite temperature. 
However, the thermal states are of more interest in 2D, where they can undergo finite temperature phase transformations.  
A thermal PEPS with ancillas was considered in Ref. \cite{Czarnik}, where a finite temperature phase diagram
of the 2D quantum Ising model in a transverse field was obtained. This approach is further developed in this
paper to a fermionic thermal PEPS, with a benchmark application to a 2D spinless Hubbard model.

Before we proceed, let us note that the PEPS with ancillas is not the only way to attack the strongly correlated thermal 
states. A very interesting alternative was developed in a series of papers \cite{ChinaT} where, instead of the imaginary 
time evolution, a tensor network representing the partition function is contracted by subsequent tensor 
renormalizations in the imaginary time and space dimensions. Yet another interesting alternative, presented in 
Ref. \cite{Poulin}, is based on linear optimization of local density matrices at finite $T$. Finally, alternative 
representations for fermionic states are also developed \cite{Ferris}.

The paper is organized as follows. In Section \ref{tPEPS} we introduce fermionic PEPS with ancillas at finite
temperature and outline the algorithm in most general terms. In brief Section \ref{Hubbard} the Hubbard model
for spinless fermions on a square lattice is introduced with a hoping term, a symmetry breaking term, and
a nearest-neighbor (NN) attraction. The following section \ref{Suzuki} introduces the imaginary time evolution
operator, its second order Suzuki-Trotter decomposition for the spinless Hubbard model, and their diagrammatic representation 
in terms of the tensor network. The PEPS tensors require renormalization/truncation of their bond indices after every
Suzuki-Trotter gate. The renormalization procedure is described in Section \ref{renormalization}, where it is illustrated
with a series of diagrams. It is a variation on the corner matrix renormalization \cite{CMR}.
In Section \ref{results} we report benchmark results of the algorithm in the spinless Hubbard model. 
We conclude in Section \ref{summary}.  

%%%%%%%%%%%%%%%%%%%%%%%%%%%%%%%%%%%%%%%%%%%%%%%%%%%%%%%%%%%%%%%%%%%%%%%%%% 
\section{ PEPS at finite temperature}\label{tPEPS}
%%%%%%%%%%%%%%%%%%%%%%%%%%%%%%%%%%%%%%%%%%%%%%%%%%%%%%%%%%%%%%%%%%%%%%%%%%
 
We consider spinless fermions on an infinite square lattice with a Hamiltonian ${\cal H}$. 
Every site has two Fock states numbered by their fermionic occupation number $i=0,1$. 
Every site is accompanied by a fermionic ancilla with Fock states $a=0,1$.
The enlarged Hilbert space is spanned by Fock states $\prod_s |a_s,i_s\rangle$, 
where the ordered product runs over all lattice sites $s$. 
The infinite temperature state
$
\rho(\beta=0) \propto {\bf 1}
$
is obtained from a pure state in the enlarged space by tracing out the ancillas, 
\be 
\rho(0) ~=~ {\rm Tr}_{\rm a}|\psi(0)\rangle\langle\psi(0)|~,
\ee
where 
\be 
|\psi(0)\rangle ~=~ \prod_s \frac{1}{\sqrt2} \left( |0_s,0_s\rangle+|1_s,1_s\rangle \right)~
\label{psi0}
\ee
is a product of maximally entangled states of every site with its ancilla. 
The state $\rho(\beta)\propto e^{-\beta {\cal H}}$ at a finite $\beta$ is obtained from
\be 
|\psi(\beta)\rangle~\propto~
e^{-\frac12{\cal H}\beta}~|\psi(0)\rangle~\equiv~
U(\beta)~|\psi(0)\rangle~
\ee
after an imaginary time evolution.

\begin{figure}[t]
\vspace{0.3cm}
\includegraphics[width=1.0\columnwidth,clip=true]{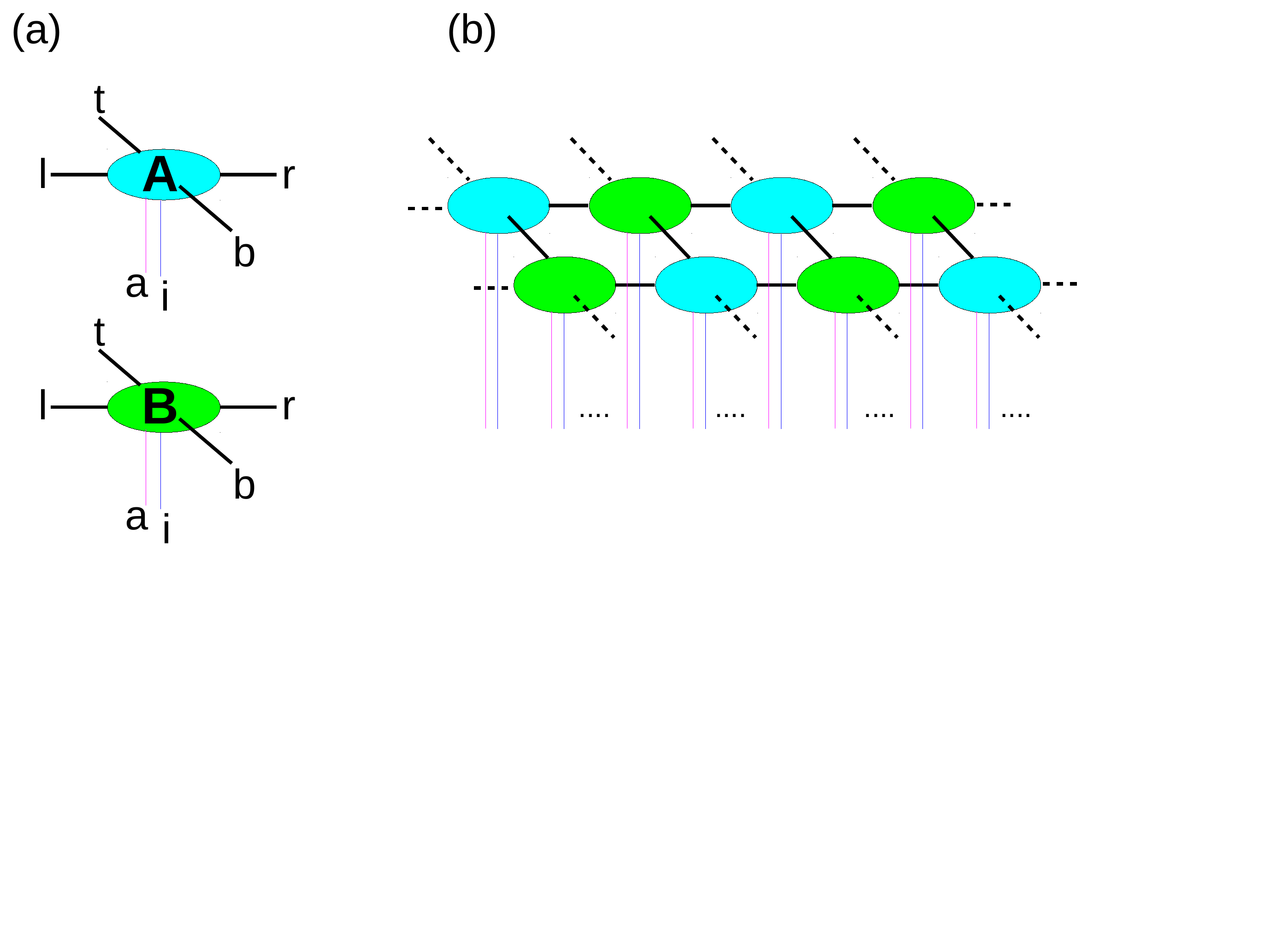}
\vspace{-3.3cm}
\caption{ 
In (a), a graphic representation of the tensors $A^{ia}_{ltrb}$ and $B^{ia}_{ltrb}$.
In (b), the amplitude $\Psi_{A,B}[\{i,a\}]$ with all bond indices connecting nearest-neighbors tensors contracted.
The index contraction is represented by a line connecting two tensors. 
Each crossing between two lines implies a fermionic swap factor:
when the indices on both lines are odd, 
then their contribution to the contraction is multiplied by $-1$.
The inclusion of this swap gate follows the strategy proposed
by Corboz {\it et al.} in Ref. \cite{fermions}.     
The tensors make a checkerboard lattice with sublattices $A$ and $B$. 
The ansatz becomes translationally invariant when the tensors $A=B$ 
(modulo a gauge transformation on the bond index).
}
\label{FigPeps}
\end{figure}

For an efficient simulation of the evolution we represent $|\psi(\beta)\rangle$ by a PEPS on the checkerboard lattice, see Fig. \ref{FigPeps}. 
The lattice has two sublattices, $A$ and $B$, 
with the same tensor, $A^{ia}_{ltrb}(\beta)$ and $B^{ia}_{ltrb}(\beta)$ respectively, at every site of the sublattice.
Here $i$ and $a$ are fermion and ancilla indices respectively, 
and $l,t,r,b=0,...,D-1$ are bond indices to contract the tensor with its nearest neighbors.
The tensors are parity preserving:
only the tensor elements with an even sum $i+a+l+t+r+b$ can be non-zero.
By construction, 
the ansatz is translationally invariant with respect to diagonal moves,
but when the tensors $A$ and $B$ are the same (up to a gauge transformation on bond indices) 
it gains full translational invariance.
The state is
\be 
|\psi(\beta)\rangle ~=~  \sum_{\{a_s,i_s\}} ~ \Psi_{A,B}[\{a_s,i_s\}] ~ \prod_s|a_s,i_s\rangle~.
\ee
Here the sum runs over all indices $a_s,i_s$ at all sites. 
The amplitude $\Psi_{A,B}[\{a,i\}]$ is the tensor contraction in Fig. \ref{FigPeps}b. 
The initial product state (\ref{psi0}) can be represented by 
\be 
A^{ia}_{ltrb}~=~ 
B^{ia}_{ltrb}~=~
\delta^{ia} ~ \delta_{l0} ~ \delta_{t0} ~ \delta_{r0} ~ \delta_{b0}~
\label{A0}
\ee
with the minimal bond dimension $D=1$.

%%%%%%%%%%%%%%%%%%%%%%%%%%%%%%%%%%%%%%%%%%%%%%%%%%%%%%%%%%%%%%%%%%%%%%%%%%%%
%%%%%%%%%%%%%%%%%%%%%%%%%%%%%%%%%%%%%%%%%%%%%%%%%%%%%%%%%%%%%%%%%%%%%%%%%%%%

%%%%%%%%%%%%%%%%%%%%%%%%%%%%%%%%%%%%%%%%%%%%%%%%%%%%%%%%%%%%%%%%%%%%%%%%%% 
\section{ Spinless Hubbard model }\label{Hubbard}
%%%%%%%%%%%%%%%%%%%%%%%%%%%%%%%%%%%%%%%%%%%%%%%%%%%%%%%%%%%%%%%%%%%%%%%%%%

We proceed with the spinless Hubbard Hamiltonian
\bea 
{\cal H} &=& 
- \sum_{\langle s_A,s_B\rangle} c_{s_A}^\dag c_{s_B} + {\rm H.c.} + \nonumber\\      
         & & 
- \delta \sum_{\langle s_A,s_B\rangle} c_{s_A} c_{s_B} + {\rm H.c.} + \nonumber\\       
         & &
- g \sum_{\langle s_A,s_B\rangle} \left(n_{s_A}-\frac12\right)\left(n_{s_B}-\frac12\right)           ~ \nonumber\\
         &\equiv&
{\cal H}_{B\to A} + 
{\cal H}_{A\to B} + 
{\cal H}_{\delta} +
{\cal H}_{\delta}^\dag + 
{\cal H}_{g}.
\label{calH}
\eea
Here the index $s_A$ ($s_B$) runs over the sublattice $A$ ($B$), $c_s$ is a fermionic annihilation operator, and $n_s=c^\dag_s c_s$ 
is the occupation number. The $\delta$-term is an explicit $U(1)$ symmetry breaking $p$-wave pairing, and the $g$-term is 
a nearest-neighbor attraction.

%%%%%%%%%%%%%%%%%%%%%%%%%%%%%%%%%%%%%%%%%%%%%%%%%%%%%%%%%%%%%%%%%%%%%%%%%% 
\section{ Suzuki-Trotter decomposition }\label{Suzuki}
%%%%%%%%%%%%%%%%%%%%%%%%%%%%%%%%%%%%%%%%%%%%%%%%%%%%%%%%%%%%%%%%%%%%%%%%%%

We define elementary infinitesimal evolution operators:
\bea 
U_{A\to B}(d\beta) & \equiv & 
e^{-\frac12d\beta~ {\cal H}_{A\to B}} = 
\prod_{\langle s_A,s_B \rangle} 
\left( 1+\frac12 d\beta~ c_{s_B}^\dag c_{s_A} \right),\nonumber\\
U_{B\to A}(d\beta) & \equiv & 
e^{-\frac12d\beta~ {\cal H}_{B\to A}} =
\prod_{\langle s_A,s_B \rangle} 
\left( 1+\frac12 d\beta~ c_{s_A}^\dag c_{s_B} \right),\nonumber\\
U_{\delta}(d\beta) & \equiv & 
e^{-\frac12d\beta~ {\cal H}_{\delta}} = 
\prod_{\langle s_A,s_B \rangle} 
\left( 1+\frac12 d\beta~\delta~ c_{s_A} c_{s_B} \right),\nonumber\\
U_{g}(d\beta)      & \equiv  & e^{-\frac12d\beta~ {\cal H}_{g}} \nonumber\\
                   & \propto &
\prod_{\langle s_A,s_B \rangle}
\left[ 1 + \epsilon \left(n_{s_A}-\frac12\right)\left(n_{s_B}-\frac12\right) \right],
\eea
where $\epsilon=4\tanh\left(\frac14g\frac12d\beta\right)$. Each of the evolution operators is a product
of $2$-site gates. They are the building blocks for the second order Suzuki-Trotter decomposition.

%%%%%%%%%%%%%%%%%%%%%%%%%%%%%%%%%%%%%%%%%%%%%%%%%%%%%%%%%%%%%%%%%%%%%%%%%%%%
\begin{figure}[t!]
\vspace{-0.3cm}
\includegraphics[width=1.0\columnwidth,clip=true]{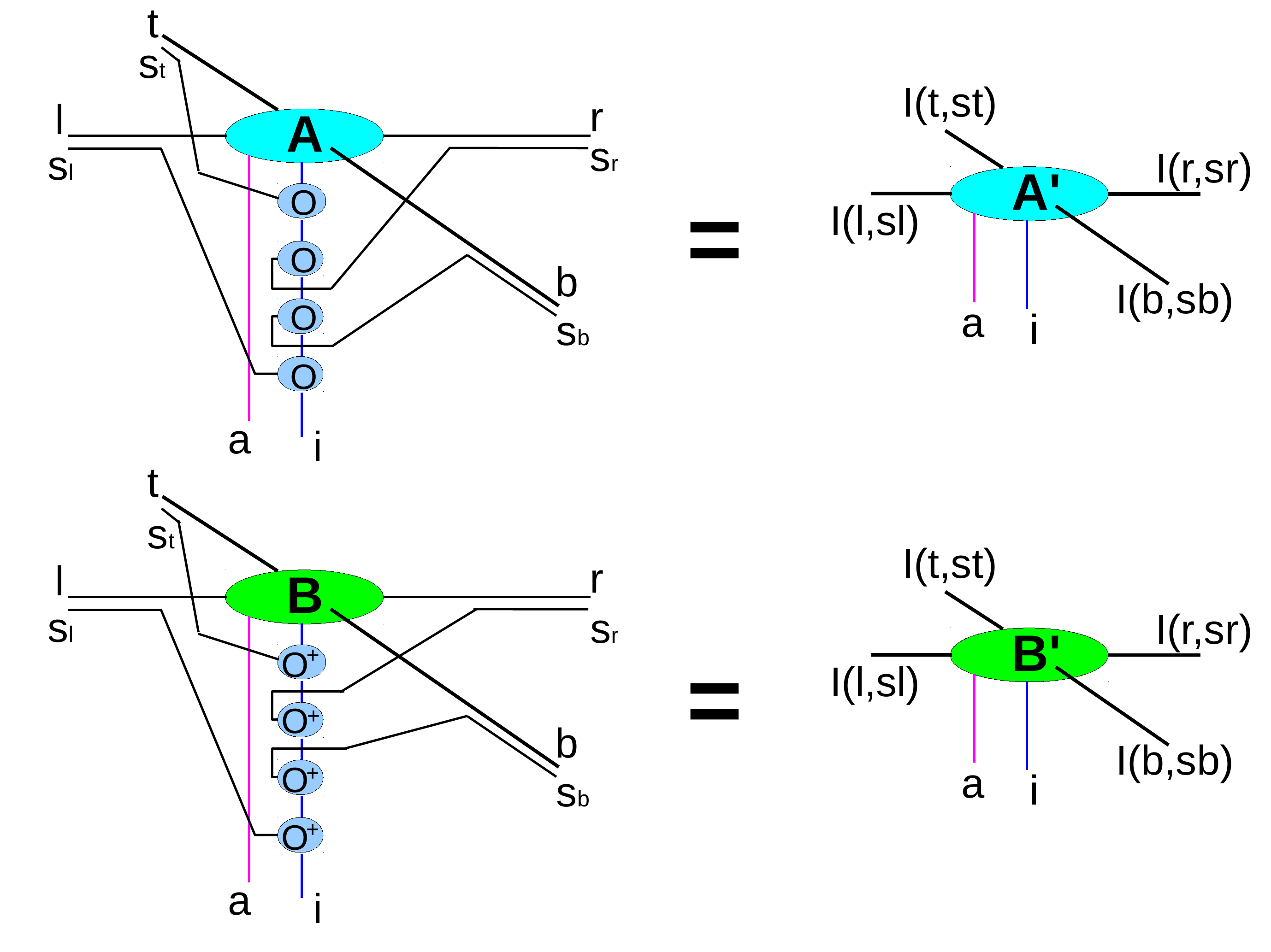}
\vspace{-0.5cm}
\caption{ 
Tensor contractions representing the action of $U_{A\to B}(d\beta)$ on the tensors $A$ and $B$, 
compare with Eqs. (\ref{A},\ref{B}). 
Here each crossing between two lines implies a fermionic swap factor $S$.
The operator $o=1$ when its bond index $s_x=0$, or $o=c\sqrt{d\beta/2}$ when $s_x=1$.
Notice that the tensor $o$ is parity preserving: the sum of its three indices is even. 
Notice also that when more than one index $s_x=1$ the contraction is zero: 
a creation/annihilation operator cannot be applied more than once.
}
\label{FigAprime}
\end{figure}
%%%%%%%%%%%%%%%%%%%%%%%%%%%%%%%%%%%%%%%%%%%%%%%%%%%%%%%%%%%%%%%%%%%%%%%%%%%%

The action of the hopping operator, say, $U_{A\to B}(d\beta)$ maps the tensors $A$ and $B$ to new tensors:
\bea
&& 
\left(A'\right)^{ia}_{I(l,s_l),I(t,s_t),I(r,s_r),I(b,s_b)} ~=~ \label{A}\\
&&
\delta_{s,0}~
A^{ia}_{ltrb}+ \nonumber\\
&&
\delta_{s,1}~
\sqrt{\frac{d\beta}{2}}~
S_{s_tl}S_{s_ta}S_{s_la}S_{s_rb}~
%(-1)^{s_r+s_b}
\sum_j
\langle i |c| j \rangle 
A^{ja}_{ltrb}~, \nonumber\\
&& 
\left(B'\right)^{ia}_{I(l,s_l),I(t,s_t),I(r,s_r),I(b,s_b)} ~=~ \label{B}\\
&&
\delta_{s,0}~
B^{ia}_{ltrb}+ \nonumber\\
&&
\delta_{s,1}~
\sqrt{\frac{d\beta}{2}}~
S_{s_tl}S_{s_ta}S_{s_la}S_{s_rb}~
\sum_j
\langle i |c^\dag| j\rangle 
B^{ja}_{ltrb}~, \nonumber
\eea
see the diagrams in Figure \ref{FigAprime}.
Here the gate indices $s_l,s_t,s_r,s_b\in\left\{0,1\right\}$ and $s=s_l+s_t+s_r+s_b$. 
An odd $s_x=1$ means a transfer of one fermion along the bond $x$. 
$I(x,s_x)$ is an invertible parity-preserving index function.
$S$ is a fermionic swap factor: $S_{ab}=-1$ when both $a$ and $b$ are odd and $+1$ otherwise.
Equations (\ref{A},\ref{B}) are an exact map, 
but the new tensors $A'$ and $B'$ have the bond dimension $2D$ instead of the original $D$. 
  
\begin{figure}[t]
\vspace{-0.3cm}
\includegraphics[width=1.0\columnwidth,clip=true]{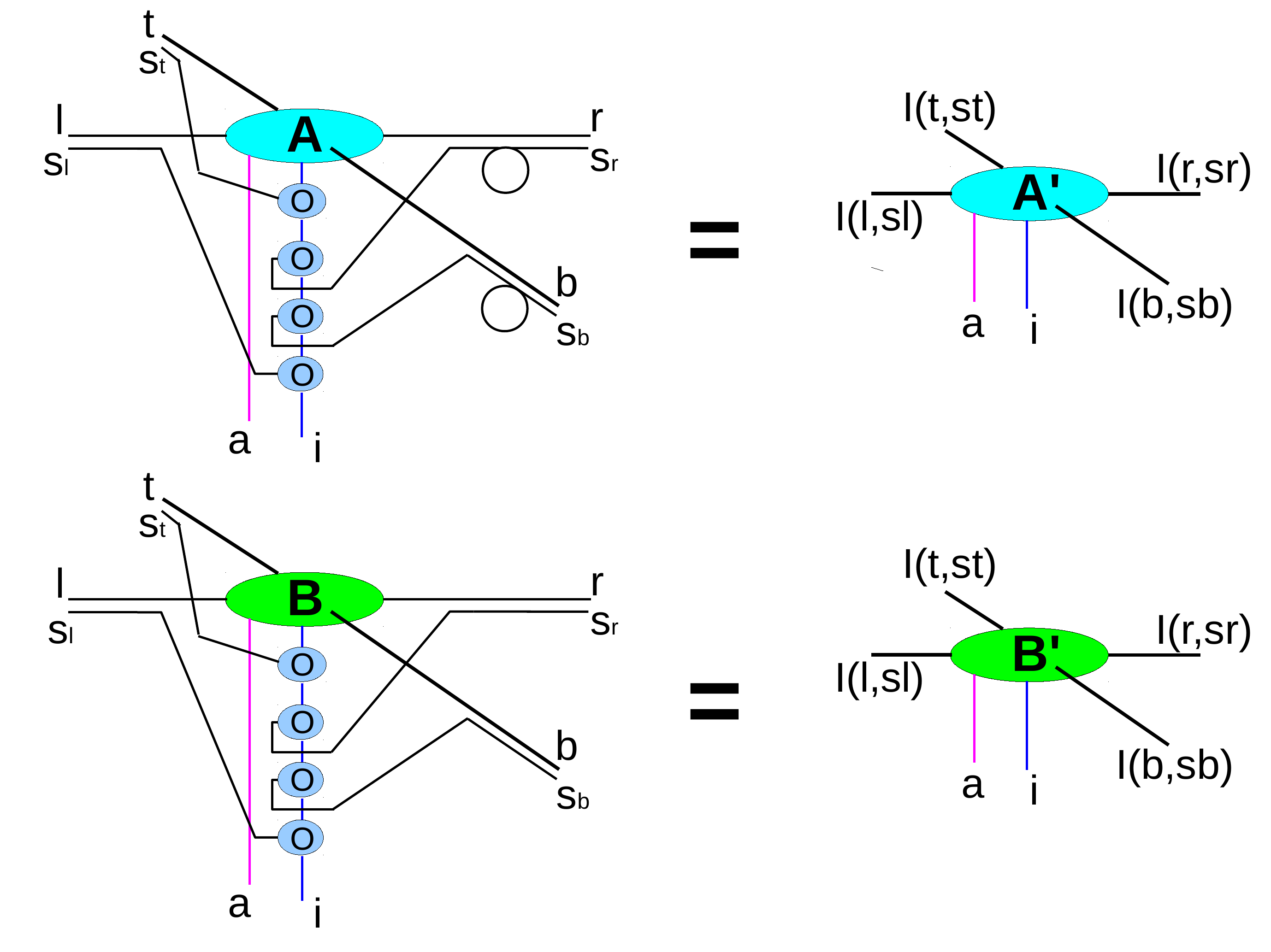}
\vspace{-0.5cm}
\caption{ 
The tensor contraction representing the action of $U_{\delta}(d\beta)$ on the tensors $A$ and $B$,
see Eqs. (\ref{Adelta},\ref{Bdelta}).
The factors $(-1)^{s_r}=S_{s_rs_r}$ and $(-1)^{s_b}=S_{s_bs_b}$ are represented here by the self-crossing loops.
}
\label{FigAdelta}
\end{figure}

The same is true for the action of $U_\delta$, 
see the diagrams in Figure \ref{FigAdelta},
\bea
&& 
\left(A'\right)^{ia}_{I(l,s_l),I(t,s_t),I(r,s_r),I(b,s_b)} ~=~ \label{Adelta}\\
&&
\delta_{s,0}~
A^{ia}_{ltrb}+ \nonumber\\
&&
\delta_{s,1}~
\sqrt{\delta\frac{d\beta}{2}}~
S_{s_tl}S_{s_ta}S_{s_la}S_{s_rb}~
\sum_j
\langle i |c| j \rangle 
A^{ja}_{ltrb}~, \nonumber\\
&& 
\left(B'\right)^{ia}_{I(l,s_l),I(t,s_t),I(r,s_r),I(b,s_b)} ~=~ \label{Bdelta}\\
&&
\delta_{s,0}~
B^{ia}_{ltrb}+ \nonumber\\
&&
\delta_{s,1}~
\sqrt{\delta\frac{d\beta}{2}}~
S_{s_tl}S_{s_ta}S_{s_la}S_{s_rb}~
(-1)^{s_b+s_r}
\sum_j
\langle i |c| j\rangle 
B^{ja}_{ltrb}~, \nonumber
\eea
just as for the action of $U_g$, the same for both $A$ and $B$,
\be
\left(A'\right)^{ia}_{I(l,s_l),I(t,s_t),I(r,s_r),I(b,s_b)}=
\epsilon^{s/4}
\left(i-\frac12\right)^{s/2} 
A^{ia}_{ltrb}. \label{Ag}
\ee
Since $U_g$ does not transfer fermions between different sites, in Eq. (\ref{Ag}) the gate indices $s_x\in\{0,2\}$ are even 
and $I(x,s_x)$ is preserving the parity of $x$. 

%%%%%%%%%%%%%%%%%%%%%%%%%%%%%%%%%%%%%%%%%%%%%%%%%%%%%%%%%%%%%%%%%%%%%%%%%%%%
%%%%%%%%%%%%%%%%%%%%%%%%%%%%%%%%%%%%%%%%%%%%%%%%%%%%%%%%%%%%%%%%%%%%%%%%%%%%

%%%%%%%%%%%%%%%%%%%%%%%%%%%%%%%%%%%%%%%%%%%%%%%%%%%%%%%%%%%%%%%%%%%%%%%%%% 
\section{ Renormalization of bond indices }\label{renormalization}
%%%%%%%%%%%%%%%%%%%%%%%%%%%%%%%%%%%%%%%%%%%%%%%%%%%%%%%%%%%%%%%%%%%%%%%%%% 

After every gate $U$, the new bond dimension $2D$ has to be truncated back to $D$ 
in a way least distortive to the new PEPS $|\psi'\rangle$ build out of $A'$ and $B'$. 
This is done by an application of isometries $w$ that map from $2D$ to $D$ dimensions:
\be 
\sum_{l',t',r',b'}
w_l^{l'}~
w_t^{t'}~
w_r^{r'}~
w_b^{b'}~
\left(A'\right)^{ia}_{l't'r'b'}~=~
{\rm (new)}~
A^{ia}_{ltrb}~,
\ee 
see Fig. \ref{FigRenB}d. 
$w$ must be parity preserving for the new $A$ to preserve parity. 
The isometries should be optimized 
to be the least destructive to the norm squared $\langle\psi'|\psi'\rangle$. 

\begin{figure}[t!]
\vspace{-0.3cm}
\includegraphics[width=1.0\columnwidth,clip=true]{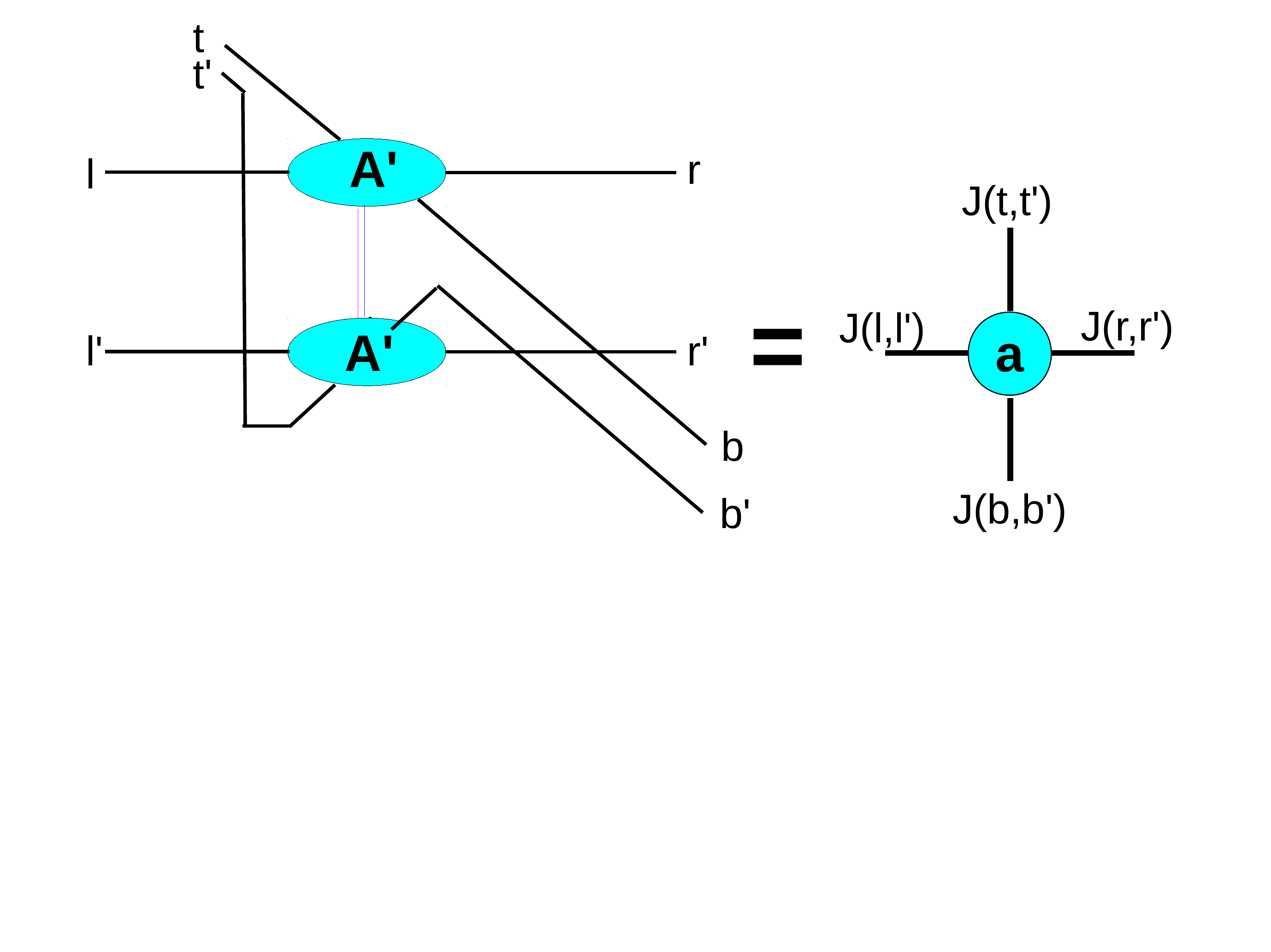}
\vspace{-3.5cm}
\caption{ 
A contraction of the new tensor $A'$ with its conjugate that makes a transfer matrix $a$.
Here $J(x,x')$ is an invertible parity-preserving index function.
}
\label{FigAA}
\end{figure}

\begin{figure}[t!]
\vspace{0.0cm}
\includegraphics[width=1.0\columnwidth,clip=true]{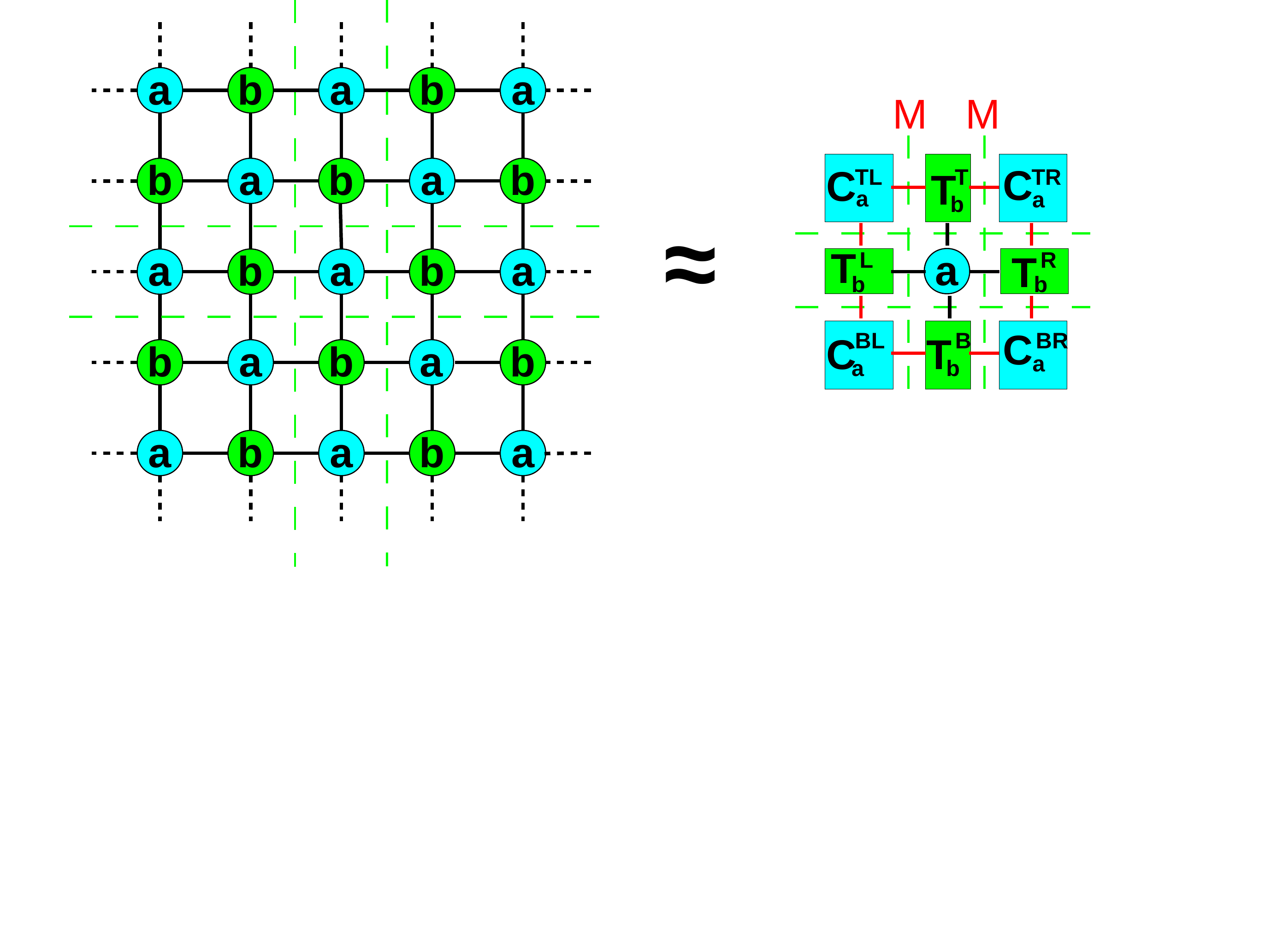}
\vspace{-3.5cm}
\caption{ 
The contraction of the transfer matrices on the LHS is the norm squared ${\rm Tr}~\rho(\beta)=\langle\psi(\beta)|\psi(\beta)\rangle$.
This contraction cannot be done exactly,
hence it is approximated by the contraction on the RHS with the corner matrices $C$ and the transfer matrices $T$. 
Their (red) environmental indices have an environmental bond dimension $M$. 
The parity-preserving $C$'s  and $T$'s should be optimized so that 
to the transfer matrix $a$ in the center 
its environment on the RHS should appear the same as its exact environment on the LHS 
as much as possible. 
Their iterative construction is described in Fig. \ref{FigRenC}.
}
\label{FigCVH}
\end{figure}

\begin{figure}[t!]
\vspace{0.0cm}
\includegraphics[width=1.0\columnwidth,clip=true]{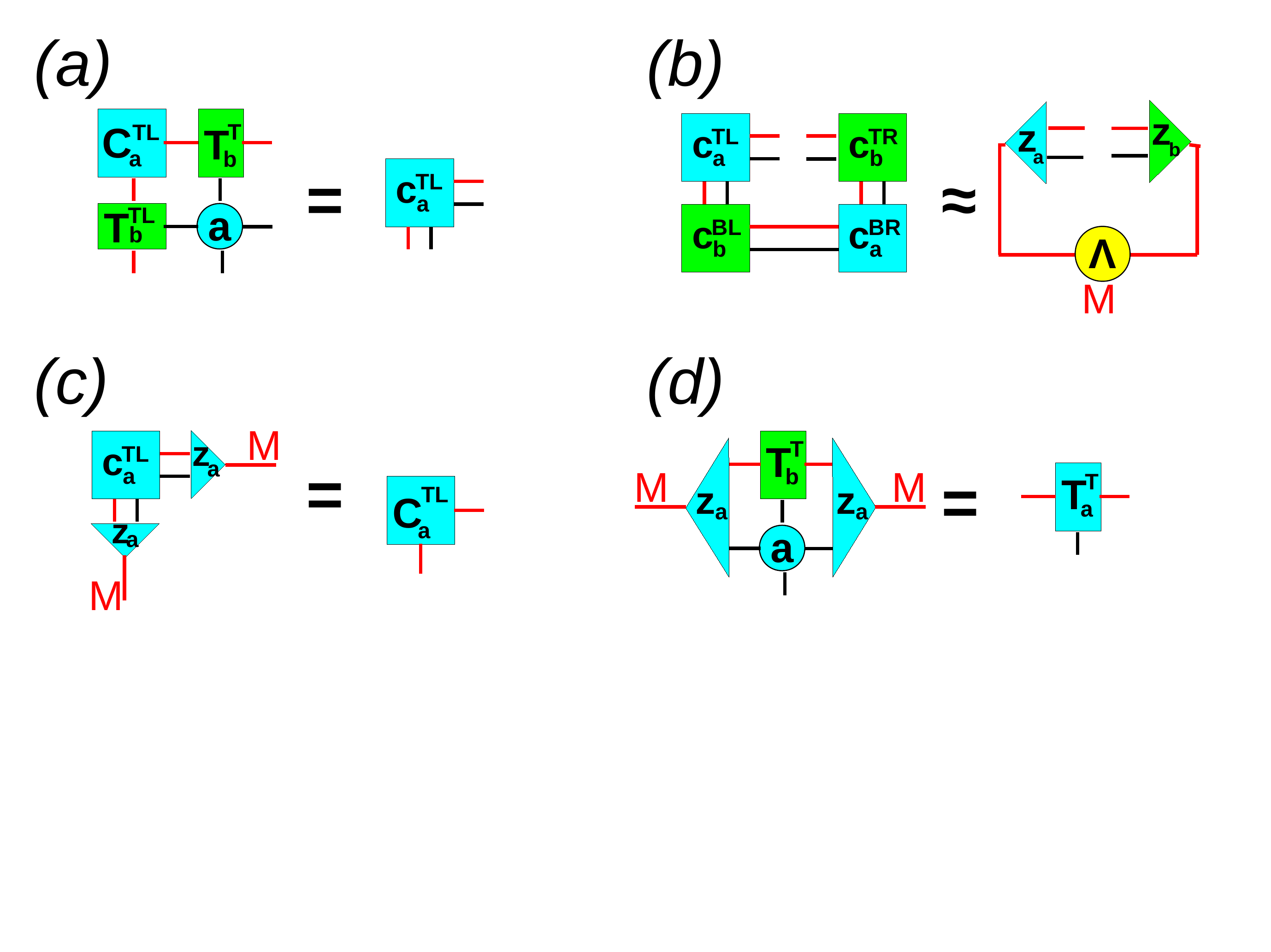}
\vspace{-3.0cm}
\caption{ 
The optimal environmental tensors $C,T$ are obtained by repeating a renormalization procedure until convergence. 
The procedure has four steps.
In step (a), 
the corner matrices $C$ are expanded to include the adjacent $T$-tensors and transfer matrices $a$ or $b$.
In step (b), 
the expanded corner matrices $c$ are contracted to form a matrix 
whose singular value decomposition is truncated to $M$ leading singular values $\Lambda$.
The corresponding left and right singular vectors define the parity-preserving isometries $z_a$ and $z_b$. 
In step (c),
the isometries are used to renormalize the expanded corner matrices $c$ 
and make new corner matrices $C$.
In step (d),
the same isometries are used to renormalize expanded $T$-tensors
and make new tensors $T$.
The four-step procedure is repeated until convergence of the singular values.
}
\label{FigRenC}
\end{figure}

\begin{figure}[t!]
\vspace{0.0cm}
\includegraphics[width=0.99\columnwidth,clip=true]{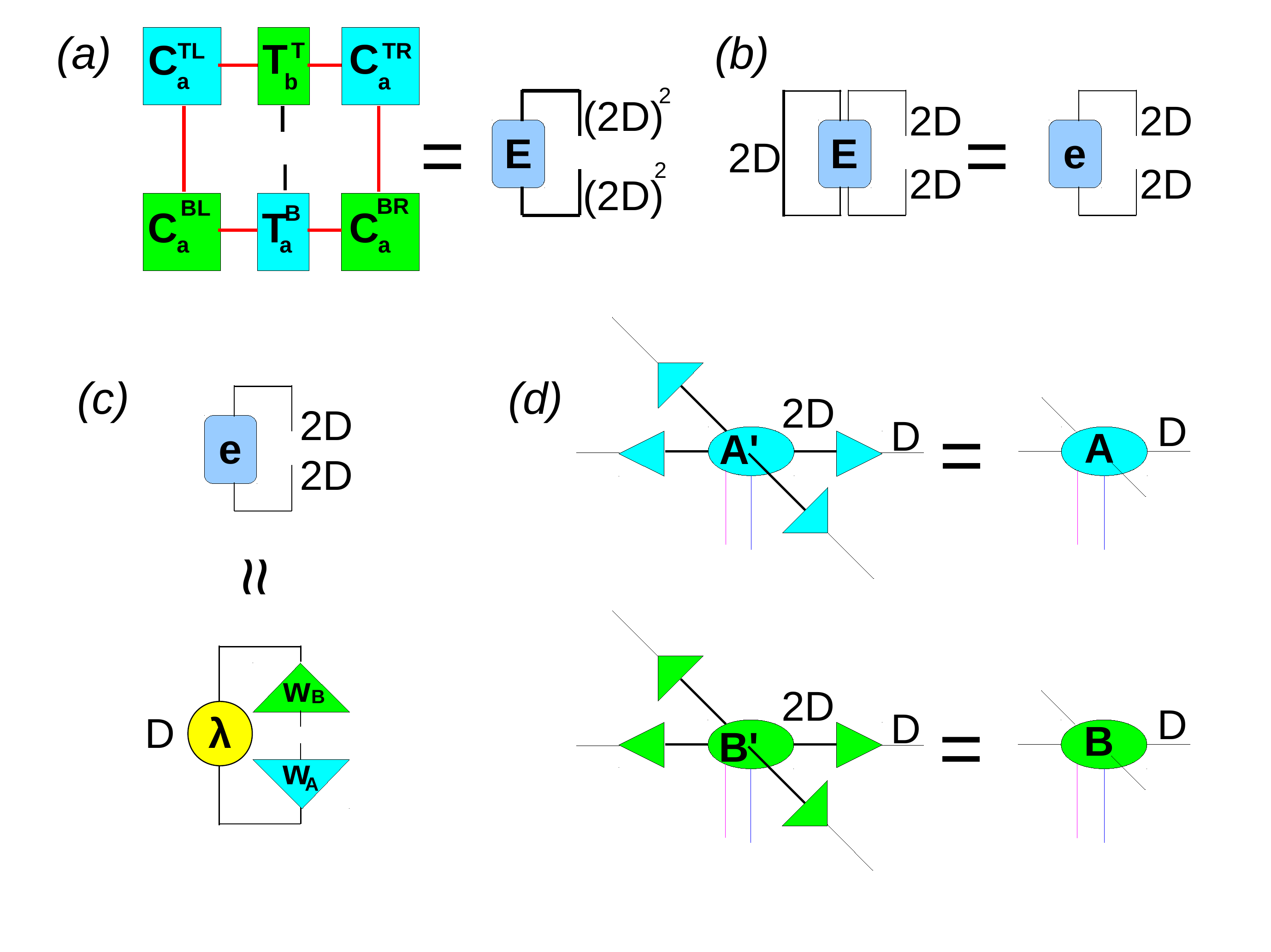}
\vspace{-1.0cm}
\caption{ 
Once the environmental tensors $C,T$ have been converged, 
one can renormalize the bond indices of the new PEPS tensors $A'$ and $B'$.
The renormalization proceeds in four steps.
In (a),
the diagram would be an approximate representation of the norm squared of the state,
if not for the one uncontracted bond in the center.
The diagram is an environment $E$ for the uncontracted bond.
In (b),
each of the two indices of $E$ can be represented by two indices of dimension $2D$ in such a way that 
the right/left index corresponds to the top/bottom tensor $A'$ or $B'$ in Fig. \ref{FigAA}. 
After the left indices of $E$ are traced out, 
one obtains an environment $e$ for an uncontracted bond in the top PEPS layer.
In (c),
the environment $e$ is subject to a singular value decomposition 
that is truncated to the $D$ leading singular values $\lambda$.
Their corresponding singular vectors define the isometries $w_A$ and $w_B$.
In (d),
the isometries renormalize the tensors $A'$ and $B'$ back to new tensors $A$ and $B$ with the original bond dimension $D$.
This renormalization completes the action of the evolution gate $U$. 
}
\label{FigRenB}
\end{figure}

The construction of the best isometry described in Figs. \ref{FigAA},\ref{FigCVH},\ref{FigRenC},\ref{FigRenB} 
is a variant of the corner matrix renormalization \cite{CMR}.
It requires calculation of tensor environments for $A'$ and $B'$ in the network representing the norm squared $\langle\psi'|\psi'\rangle$.
This environment cannot be calculated exactly in an efficient way. 
This is why it is replaced by an effective environment, 
made of the environmental tensors $C$ and $T$,
that should appear to the tensors $A'$ and $B'$ as close to the exact one as possible.
The environmental tensors are contracted with each other by the environmental indices of dimension $M$.   
Increasing $M$ should make the effective environment more accurate. 
%The overall cost of renormalizing $B$ back to the bond dimension $D$ is polynomial in both $D$ and $M$. 
%It is dominated by the calculation of ?? in Fig. ??  that scales like $M^?D^?$. 

At the beginning of the time evolution the environmental tensors $C,T$ are initialized with random numbers.
After every time step we add weak noise to the converged tensors before they are re-used in the next time step. 
The noise prevents the tensors from being trapped in a subspace of a lower dimension $D$ or $M$.
A bit more technical issue concerns the construction of the new $C$ and $T$ in Figs. \ref{FigRenC}c and \ref{FigRenC}d respectively. 
In principle, all $M$ leading singular vectors $z$ can be used in this contraction, even those corresponding to singular values equal to numerical zero.
By construction, the ``zero vectors'' do not make any difference when $C,T$ are contracted as in the norm-squared of PEPS. 
However, we found the algorithm to be unstable unless we set the (numerically inaccurate) zero vectors to zero. 
The inaccuracies do make a difference when the renormalized $C,T$ are contracted in a way different than the norm-squared. 
This truncation requires a cut-off that sets the minimal singular value that is considered to be non-zero.
Its net effect is that the algorithm occasionally operates with an effective $M_{\rm eff}$ that is less than the declared $M$.

%%%%%%%%%%%%%%%%%%%%%%%%%%%%%%%%%%%%%%%%%%%%%%%%%%%%%%%%%%%%%%%%%%%%%%%%%%%%
%%%%%%%%%%%%%%%%%%%%%%%%%%%%%%%%%%%%%%%%%%%%%%%%%%%%%%%%%%%%%%%%%%%%%%%%%%%%

%%%%%%%%%%%%%%%%%%%%%%%%%%%%%%%%%%%%%%%%%%%%%%%%%%%%%%%%%%%%%%%%%%%%%%%%%%%%
%%%%%%%%%%%%%%%%%%%%%%%%%%%%%%%%%%%%%%%%%%%%%%%%%%%%%%%%%%%%%%%%%%%%%%%%%%%%

%%%%%%%%%%%%%%%%%%%%%%%%%%%%%%%%%%%%%%%%%%%%%%%%%%%%%%%%%%%%%%%%%%%%%%%%%%%%
%%%%%%%%%%%%%%%%%%%%%%%%%%%%%%%%%%%%%%%%%%%%%%%%%%%%%%%%%%%%%%%%%%%%%%%%%%%%

%%%%%%%%%%%%%%%%%%%%%%%%%%%%%%%%%%%%%%%%%%%%%%%%%%%%%%%%%%%%%%%%%%%%%%%%%%%%
%%%%%%%%%%%%%%%%%%%%%%%%%%%%%%%%%%%%%%%%%%%%%%%%%%%%%%%%%%%%%%%%%%%%%%%%%%%%

%%%%%%%%%%%%%%%%%%%%%%%%%%%%%%%%%%%%%%%%%%%%%%%%%%%%%%%%%%%%%%%%%%%%%%%%%% 
\section{ Benchmark results }\label{results}
%%%%%%%%%%%%%%%%%%%%%%%%%%%%%%%%%%%%%%%%%%%%%%%%%%%%%%%%%%%%%%%%%%%%%%%%%% 

Here we summarize results for the Hamiltonian (\ref{calH}) with or without the symmetry-breaking 
$\delta$-term or the interaction $g$-term. 
%In all Figures the imaginary time step is $d\beta=0.01$. 
%The results appear to be converged in $d\beta$.

%%%%%%%%%%%%%%%%%%%%%%%%%%%%%%%%%%%%%%%%%%%%%%%%%%%%%%%%%%%%%%%%%%%%%%%%%% 
\subsection{ Quadratic Hamiltonian with $g=0$ }\label{resultsnog}
%%%%%%%%%%%%%%%%%%%%%%%%%%%%%%%%%%%%%%%%%%%%%%%%%%%%%%%%%%%%%%%%%%%%%%%%%% 

We begin with the exactly solvable case when the quartic interaction $g=0$. 
The ground state does not satisfy the area law for entanglement, 
but this does not preclude accurate description of its high temperature properties.  

In the absence of the $U(1)$ symmetry-breaking $\delta$-term, all imaginary time evolutions preserve 
the average density $\langle n_s\rangle=0.5$ within numerical precision. With the $\delta$-term
the density is preserved with a precision of $10^{-5}$.
 
Nonzero local averages are the NN hopping term $\langle c^\dag_{s_A} c_{s_B} \rangle$ and the NN anomalous term $\langle c_{s_A} c_{s_B} \rangle$. 
Both are obtained from a NN two-site reduced density matrix $\rho_2$. 
The numerical $\rho_2$ can be compared with its exact counterpart $\rho_2^{\rm exact}$.
Their difference can be quantified by an error
\be 
\bigg\|\frac{\rho_2}{\|\rho_2\|}   -  \frac{\rho_2^{\rm exact}}{\|\rho_2^{\rm exact}\|}\bigg\|,
\ee  
where $\| A \| = \sqrt{{\rm Tr} (AA^\dag)}$,
that gives an idea about the order of magnitude of the errors of individual matrix elements. 
A more demanding test, that goes beyond the NN $\rho_2$, is a two-site correlator:
\be 
C_r=    \langle c^\dag_{x,y} c_{x+r,y} \rangle
\ee
between sites $(x,y)$ and $(x+r,y)$. 
Again, it can be compared with its exact counterpart.

\begin{figure}[t]
\vspace{0.0cm}
\includegraphics[width=0.99\columnwidth,clip=true]{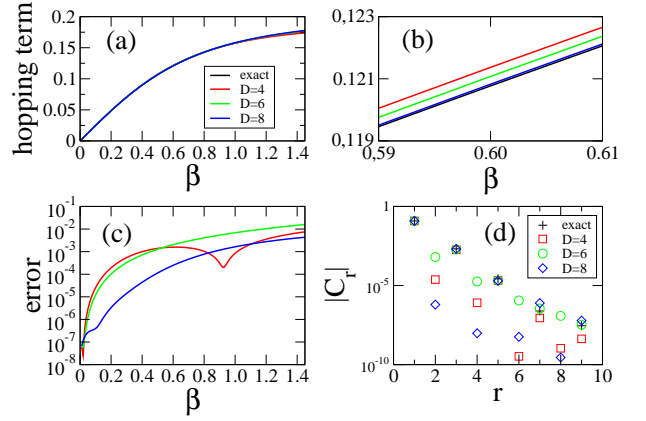}
\vspace{-0.8cm}
\caption{ 
In (a), 
the NN hopping term $\langle c^\dag_{s_A} c_{s_B} \rangle$ in function of $\beta$.
In (b), 
focus on $\beta\approx0.6$ showing convergence to the exact result with increasing bond dimension $D$.
In (c), 
the error of the two-site reduced density matrix $\rho_2$ in function of $\beta$.
In (d), 
modulus the correlation function $|C_r|$ at $\beta=0.6$. 
The exact $C_r=0$ for even $r$ and numerically it is small indeed, especially for $D=8$.
}
\label{delta0}
\end{figure}

Results for the Fermi sea ($\delta=0,g=0$) are shown in Figure \ref{delta0}. 
For the bond dimensions $D=4,6,8$ they are accurate up to $\beta=1.5$, see Fig. \ref{delta0}a. 
In this range of $D$, 
systematic convergence to the exact result with increasing $D$ is last seen around $\beta=0.6$, 
see Figs. \ref{delta0}b and \ref{delta0}c.
Figure \ref{delta0}d shows the short-ranged correlator $C_r$ at $\beta=0.6$. 
The values of $C_r$ are accurate down to $10^{-5}$. 
Apparently, 
the accuracy of the following evolution of the Fermi sea towards higher $\beta$ is limited by the bond dimension $D$.

%%%%%%%%%%%%%%%%%%%%%%%%%%%%%%%%%%%%%%%%%%%%%%%%%%%%%%%%%%%%%%%%%%%%%%%%%%%%
%%%%%%%%%%%%%%%%%%%%%%%%%%%%%%%%%%%%%%%%%%%%%%%%%%%%%%%%%%%%%%%%%%%%%%%%%%%%

\begin{figure}[t]
\vspace{-0.0cm}
\includegraphics[width=0.99\columnwidth,clip=true]{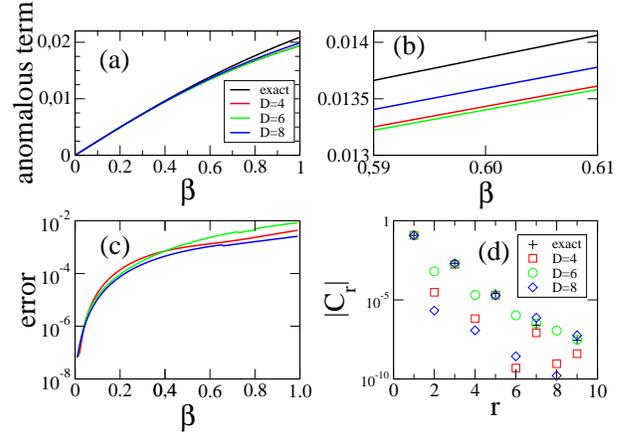}
\vspace{-0.8cm}
\caption{ 
In (a), 
the NN anomalous term $\langle c_{s_A} c_{s_B} \rangle$ in function of $\beta$.
In (b), 
focus on $\beta\approx0.6$ showing convergence to the exact result with increasing bond dimension $D$.
In (c), 
the overall error of the two-site reduced density matrix $\rho_2$.
In (d), 
modulus of the correlation function $|C_r|$ at $\beta=0.6$. 
The exact $C_r=0$ for even $r$ and numerically it is small.
}
\label{delta01}
\end{figure}

Results for noninteracting fermions with the explicit symmetry breaking $\delta$-term are shown in Figure \ref{delta01} for $\delta=0.1$. 
Up to $\beta=1.0$ the overall error of the two-site density matrix is the smallest for the largest bond dimension $D=8$, see Fig. \ref{delta01}c. 
The algorithm was not optimized to maximize accuracy of any particular observable or figure of merit, 
hence it is not quite surprising that the relative accuracy of the NN hopping term turns out better 
than the same accuracy of the NN anomalous term simply because the hopping term is much stronger than the anomalous one. 
Indeed, the error of $C_1$ in Fig. \ref{delta01}d is less than the error of the anomalous term in Fig. \ref{delta01}a. 
The values of the correlators $C_r$ in Fig. \ref{delta01}d are accurate down to $10^{-5}$. 
The exact $C_r=0$ for even $r$, and the corresponding numerical values decay to zero with increasing bond dimension $D$.

%%%%%%%%%%%%%%%%%%%%%%%%%%%%%%%%%%%%%%%%%%%%%%%%%%%%%%%%%%%%%%%%%%%%%%%%%%%%
%%%%%%%%%%%%%%%%%%%%%%%%%%%%%%%%%%%%%%%%%%%%%%%%%%%%%%%%%%%%%%%%%%%%%%%%%%%%

%%%%%%%%%%%%%%%%%%%%%%%%%%%%%%%%%%%%%%%%%%%%%%%%%%%%%%%%%%%%%%%%%%%%%%%%%% 
\subsection{ Interacting Hamiltonian }\label{resultsg}
%%%%%%%%%%%%%%%%%%%%%%%%%%%%%%%%%%%%%%%%%%%%%%%%%%%%%%%%%%%%%%%%%%%%%%%%%% 

\begin{figure}[t]
\vspace{-0.8cm}
\includegraphics[width=0.99\columnwidth,clip=true]{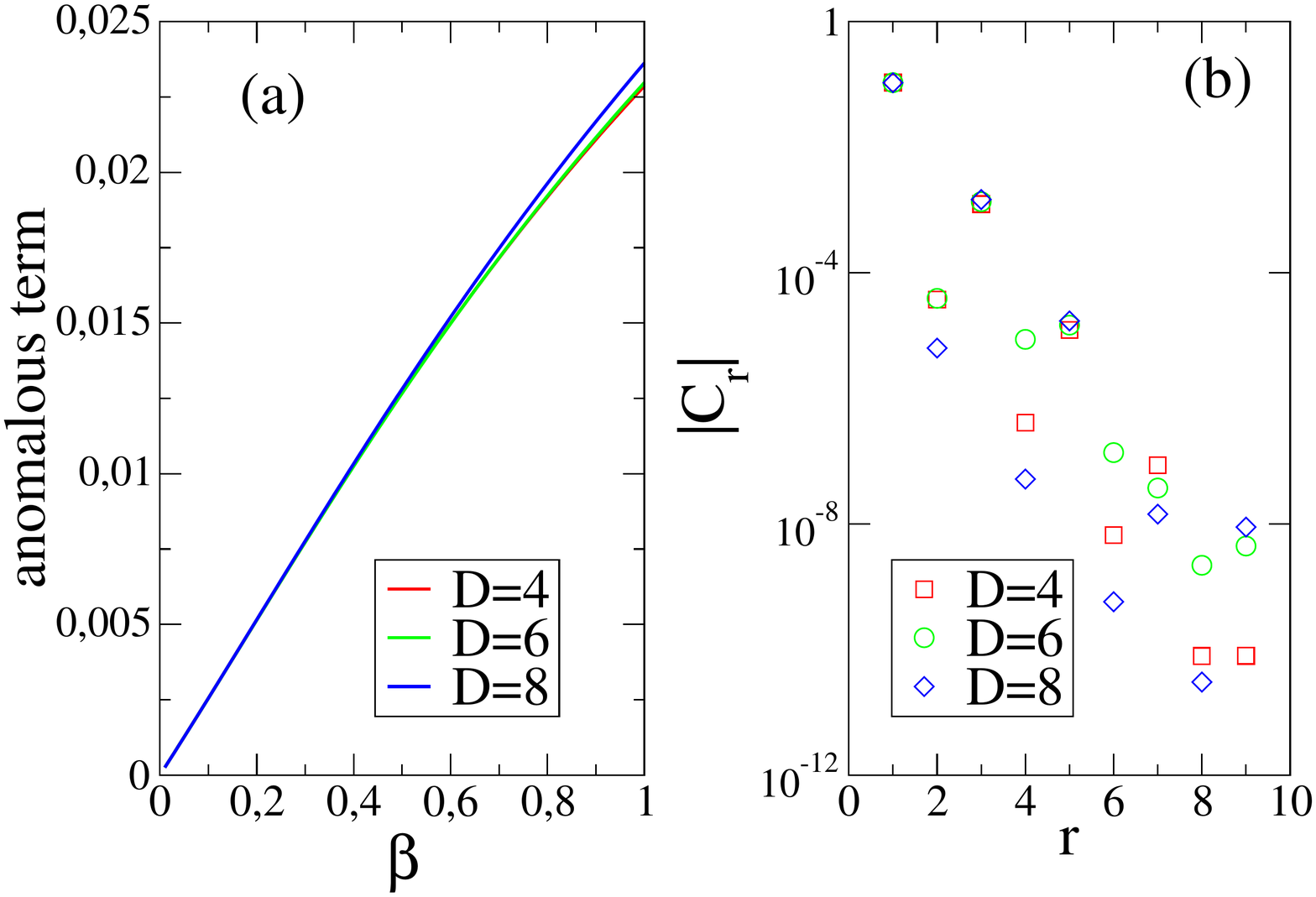}
\vspace{-0.8cm}
\caption{ 
In (a), 
the NN anomalous term $\langle c_{s_A} c_{s_B} \rangle$ in function of $\beta$.
In (b), 
modulus the correlation function $|C_r|$ at $\beta=0.6$. $|C_r|$ seems to tend to $0$ 
for even $r$.
}
\label{g1}
\end{figure}

With $g>0$ the system becomes non-integrable. 
Figure \ref{g1} shows results for a weak interaction with $g=1$ on top of $\delta=0.1$. 
The interaction enhances the $p$-wave pairing as measured by the anomalous term. 
Both the anomalous term in Fig. \ref{g1}a and the dominant correlators $C_r$ in Fig. \ref{g1}b appear converged in $D$ up to $\beta=0.6$.

%%%%%%%%%%%%%%%%%%%%%%%%%%%%%%%%%%%%%%%%%%%%%%%%%%%%%%%%%%%%%%%%%%%%%%%%%%%%
%%%%%%%%%%%%%%%%%%%%%%%%%%%%%%%%%%%%%%%%%%%%%%%%%%%%%%%%%%%%%%%%%%%%%%%%%%%%

A strong interaction $g=10$ introduces a high-temperature symmetry breaking phase transition from the symmetric phase with half-filling,
where $n=\langle n_i\rangle=1/2$, to a symmetry-broken phase where either $n=0$ or $n=1$. 
It belongs to the universality class of the 2D classical Ising model with $n-1/2$ playing the role of the order parameter.
The correlation length in the density-density correlation function 
\be
D_r=\langle n_{x+r,y} n_{x,y} \rangle-\langle n_{x+r,y} \rangle\langle n_{x,y}\rangle
\label{Dr}
\ee
diverges near the critical $\beta_c$. 
The critical correlations do not necessarily mean that the state cannot be described by a PEPS with a finite bond dimension,
but the environmental bond dimension must diverge if we want an accurate environmental tensors
either for calculation of expectation values or making a time step in the imaginary time evolution \cite{Czarnik}.
In order to smooth out the phase transition,
we add to the Hamiltonian (\ref{calH}) an explicit symmetry breaking term
\be
{\cal H}_{\rm bias} = -2 g b \sum_s n_s
\label{calHb}
\ee
with a tiny bias $b$. With a strong enough bias, it becomes possible to evolve the state across the critical regime with 
a finite environment.

\begin{figure}[t]
\vspace{+0.8cm}
\includegraphics[width=0.99\columnwidth,clip=true]{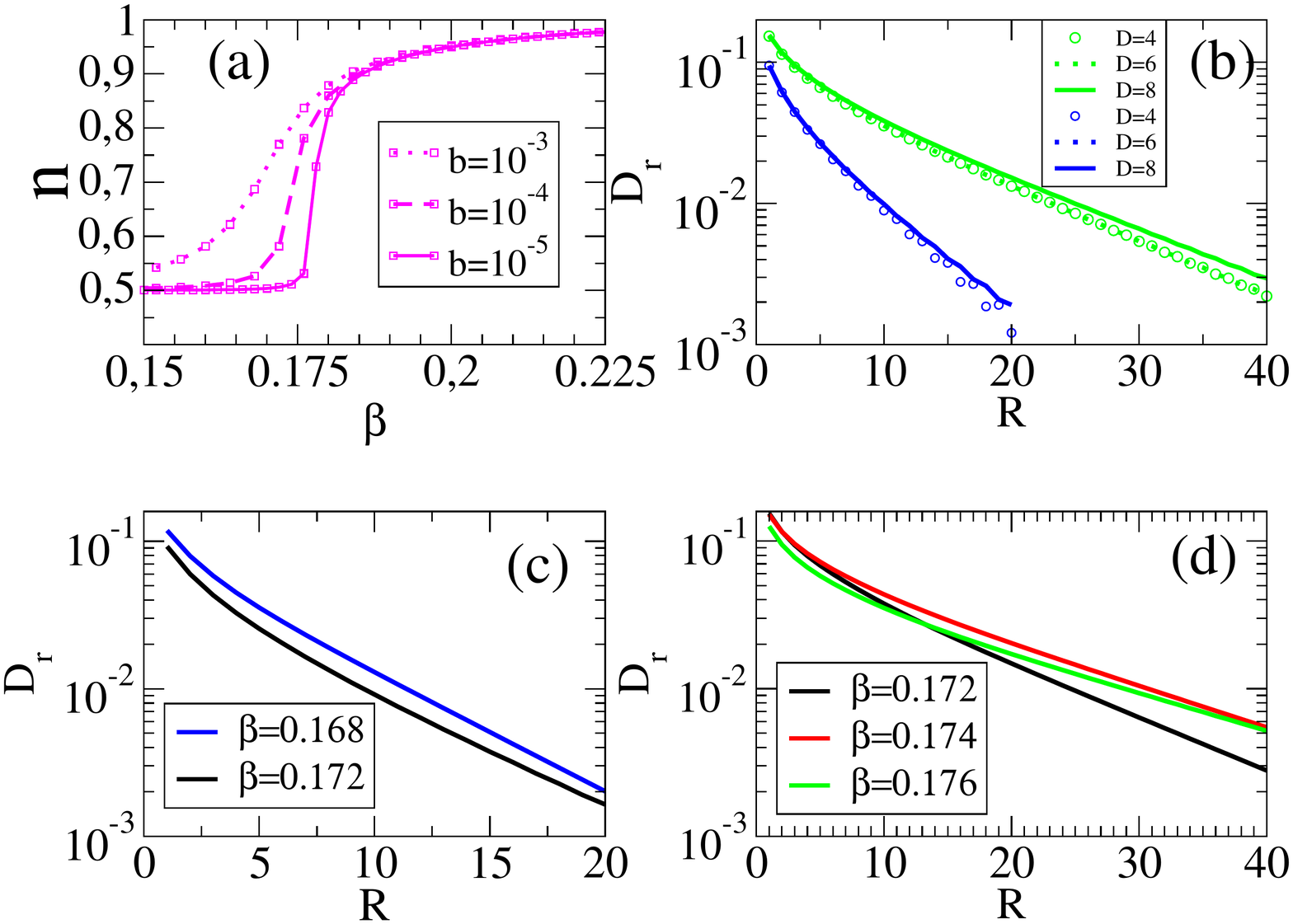}
\vspace{-0.8cm}
\caption{ 
In (a),
the average occupation number $n=\langle c_s^\dag c_s\rangle$ in function of $\beta$ for 
different values of the bias $b$. With decreasing $b$ the function develops a non-analytic
criticality.
In (b),
the connected density-density correlation function (\ref{Dr})
for $b=10^{-3},\beta=0.168$ (blue) and $b=10^{-4},\beta=0.172$ (green),
and for bond dimensions $D=4,6,8$.
In (c),
the correlator $C_R$ at the bias $b=10^{-3}$ in the critical regime where the correlation
length is the longest.
In (d),
the same as in C but for a weaker bias $b=10^{-4}$ that allows for longer correlations.
}
\label{CR}
\end{figure}

Results for $g=10$ are collected in Figure \ref{CR}. In Fig. \ref{CR}a we show the order parameter as a function of
inverse temperature for different values of the bias. Decreasing the bias results in a less analytic and more 
critical-looking curve. Figure \ref{CR}b shows the density-density correlator in the middle of the critical regime
for a bias $b=10^{-3}$ and $b=10^{-4}$. Again, the weaker bias allows for longer-range and more critical correlations. 
The plots for different bond dimensions $D=4,6,8$ demonstrate convergence with increasing $D$, 
especially at shorter distances where the correlations are more substantial. The tails of the correlation
functions are exponential, as is inevitable for a finite environmental bond dimension, but their correlation
length increases with decreasing bias. Finally, Figures \ref{CR}c and d show the correlation function for
$b=10^{-3}$ and $b=10^{-4}$ respectively at several values of $\beta$ close to $\beta_c$.

%%%%%%%%%%%%%%%%%%%%%%%%%%%%%%%%%%%%%%%%%%%%%%%%%%%%%%%%%%%%%%%%%%%%%%%%%%%%
%%%%%%%%%%%%%%%%%%%%%%%%%%%%%%%%%%%%%%%%%%%%%%%%%%%%%%%%%%%%%%%%%%%%%%%%%%%%

%%%%%%%%%%%%%%%%%%%%%%%%%%%%%%%%%%%%%%%%%%%%%%%%%%%%%%%%%%%%%%%%%%%%%%%%%% 
\section{ Summary }\label{summary}
%%%%%%%%%%%%%%%%%%%%%%%%%%%%%%%%%%%%%%%%%%%%%%%%%%%%%%%%%%%%%%%%%%%%%%%%%% 

We presented an efficient tensor network algorithm for simulation of finite temperature fermionic systems 
and its benchmark application to spinless fermions with $p$-wave pairing.
The imaginary time evolution proved to be accurate at high temperatures. 
A strong nearest-neighbor attraction introduces a high-temperature symmetry-breaking continuous phase transition.
With a tiny symmetry-breaking term the imaginary time evolution across the transition proved to be smooth,
but it still allowed for long range correlations in the critical regime.

%%%%%%%%%%%%%%%%%%%%%%%%%%%%%%%%%%%%%%%%%%%%%%%%%%%%%%%%%%%%%%%%%%%%%%%%%%%%% 
{\bf Acknowledgements. ---}
%%%%%%%%%%%%%%%%%%%%%%%%%%%%%%%%%%%%%%%%%%%%%%%%%%%%%%%%%%%%%%%%%%%%%%%%%%%%%
This work was supported in part by the Polish National Science Center (NCN) through grant 2011/01/B/ST3/00512. The research was carried out in part with the equipment purchased thanks to the financial support of the European Regional Development Fund in the framework of the Polish Innovation Economy Operational Program (contract no. POIG.02.01.00-12-023/08).
%%%%%%%%%%%%%%%%%%%%%%%%%%%%%%%%%%%%%%%%%%%%%%%%%%%%%%%%%%%%%%%%%%%%%%%%%%%%%

\end{document}